\documentclass[aip, rsi, reprint]{revtex4-1}

\usepackage[english]{babel}			 	

\usepackage{graphicx}
\usepackage{dcolumn}
\usepackage{bm}

\begin{document}
\title{Software-based discharge suppressor using an EPICS data acquisition system as a rapid prototype at the LIPAc beam extraction system}

\author{R. Ichimiya}
\affiliation{QST, Rokkasho Fusion Institute, Rokkasho, Japan}
\author{A. Jokinen}
\affiliation{F4E, Garching, Germany}
\author{A. Marqueta}
\author{B. Bolzon}
\affiliation{IFMIF/EVEDA Project Team, Rokkasho, Japan}

\date{\today}
\begin{abstract}
	Internal continuous discharge can rapidly damage high-current ion sources and
	their extraction systems composed of several electrodes at high voltage.
	To prevent this continuous discharge inside the extraction system,
	a rapid prototype using an Experimental Physics and Industrial Control System (EPICS) software system
	for data acquisition has been implemented.
	During  commissioning of the 140~mA deuterium electron cyclotron resonance ion source of 
	the Linear IFMIF Prototype Accelerator (LIPAc), discharges were often observed
	during plasma tuning of the ion source and beam optics tuning of the extraction system.
	If such continuous discharge can be avoided, discharge-related damage
	such as melting electrode edges and holes in the boron nitride
	disk in the ion source can be minimized and thus an efficient machine
	operation can be achieved.
	A veto signal is output to the machine protection system, which is then
	in charge of the RF power shutdown of the ion source for a pre-determined time. 
	The average reaction time of this system has been measured and is about 10~ms from discharge detection
	to  RF power shutdown of the ion source with a 50~Hz sampling frequency.
	This is hundreds of times slower than hardware-based implementation.
	However, it prevents almost all continuous discharges at the LIPAc ion source
	and extraction system, and is still much faster than an operator's reaction time.
\end{abstract}

\maketitle

\section{Introduction}
	The Linear IFMIF Prototype Accelerator (LIPAc) is a prototype accelerator
	to demonstrate 125~mA/9~MeV continuous-wave (CW) deuterium accelerator technologies~\cite{Knaster_LIPAc}.
	Its objective is to realize an accelerator-driven neutron
	source: the International Fusion Materials Irradiation Facility (IFMIF) that 
	provides neutron equivalent spectrum of deuterium--tritium fusion reactions
	and delivers adequate ($\mathrm{>\!\!10^{18}\, n/m^{2}/s}$) neutron flux to test materials
	to be used in future commercial fusion reactors~\cite{Knaster_IFMIF}.


	To realize this 100~mA class CW hadron accelerator, the ion source of LIPAc
	is required to produce low beam emittance (maximum value of $\mathrm{0.25 \, \pi mm \cdot mrad}$) at the radio-frequency quadrupole (RFQ) entrance with 
	well-matched Twiss parameters to minimize beam losses to less than 10\% through the RFQ~\cite{Michele}.
	These two points are to avoid damages during operation on downstream accelerator components and on the injector itself.
	A previous study showed that a boron nitride (BN) lining increases atomic ion fraction~\cite{ChalkRiver_1991}.
	Therefore, the ion source of LIPAc employs a set of BN lining disks.

\begin{figure}[!htb]
	   \centering
	   \includegraphics*[width=220pt]{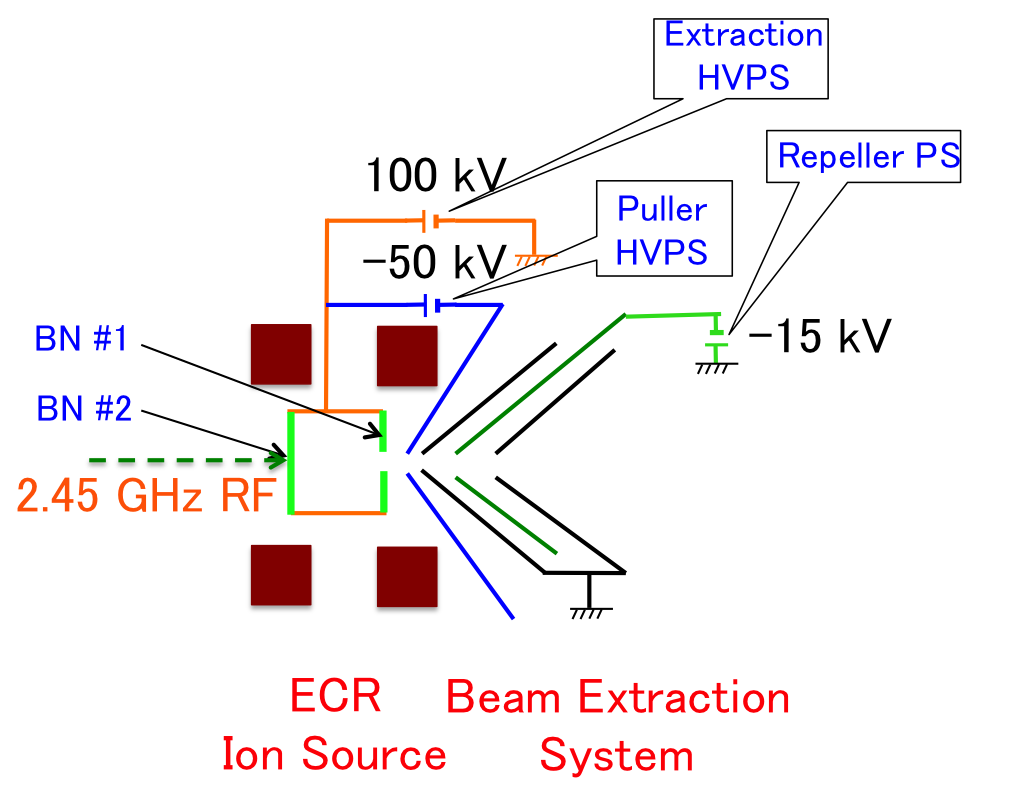}
	   \caption{LIPAc ion source and its five-electrode beam extraction system. Typical voltages for each power supply are shown.}
	   \label{Structure_IS}
	\end{figure}

	The injector of LIPAc is composed of a 2.45~GHz electron cyclotron resonance (ECR) ion source based on the CEA-Saclay SILHI source design~\cite{SILHI}
	and a low energy beam transport (LEBT) line to transport and match the beam into the RFQ using a dual solenoid
	focusing system with integrated steerers.
	Figure~\ref{Structure_IS} shows the LIPAc ECR ion source and its five-electrode beam extraction system in the accelerator column
	structure, while Table~\ref{Param_IS} shows the required beam parameters at the exit of the LIPAc injector~\cite{Gobin_LIPAc},~\cite{Delferriere_LIPAc}.
	The LIPAc ion source uses an injection of gases ($\mathrm{D_2}$ and $\mathrm{H_2}$), a 2.45 GHz 1000 W magnetron 
	RF generator, and two solenoids to generate plasma. 
	To obtain optimum beam optics, a puller electrode is used in addition to a plasma electrode.
	A negative potential electrode (called a repeller) is inserted between two ground electrodes to prevent electron
	invasion into the ion source.

	\begin{table}[hbt]
	\caption{\label{Param_IS}Beam parameters required at the exit of the LIPAc injector.}
	    \begin{ruledtabular}
	    \begin{tabular}{lc}
	       Particle type		& D+ 					\\ 
	       D+ fraction		& 99\%					\\
	       Beam energy		& 100 keV 				\\
	       Beam current		& 140 mA 				\\
	       Beam current noise	& 1\% rms				\\
	       Duty factor		& CW 					\\ 
	       				& (pulse for commissioning tests)	\\ 
	       Normalized rms transverse emittance	& $\mathrm{\leq 0.25 \pi \*mm \cdot mrad}$	\\
	   \end{tabular}
	   \end{ruledtabular}
	\end{table}
	
\section{Discharge in the Beam Extraction System}
  \subsection{Discharge phenomena and consequences}
\begin{figure}[!htb]
	   \centering
	   \includegraphics*[width=160pt]{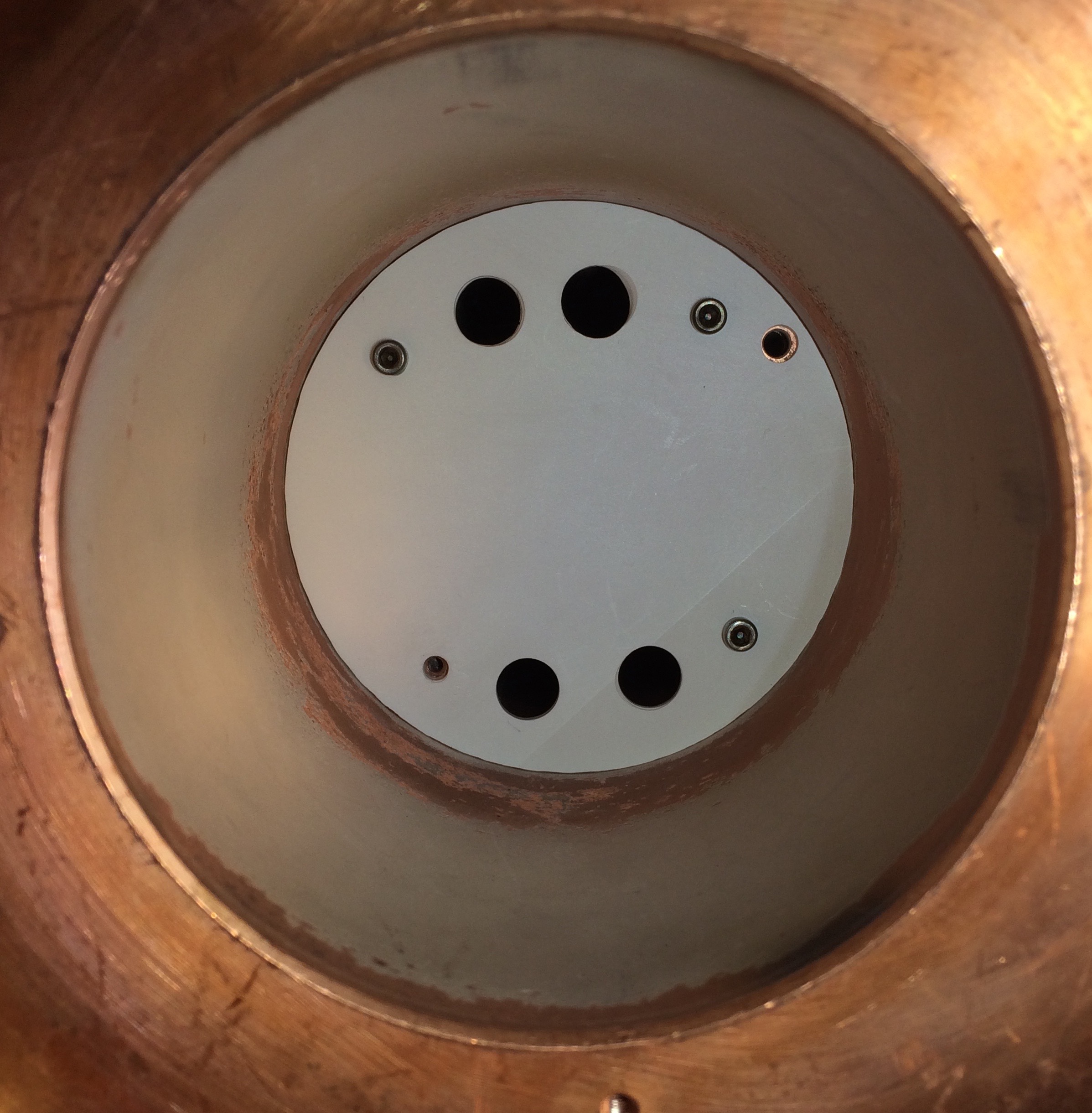}
	   \caption{A new BN disk (BN \#2 in Figure~\ref{Structure_IS}) in the LIPAc ion source.}
	   \label{IS_BN1}
	\end{figure}

\begin{figure}[!htb]
	   \centering
	   \includegraphics*[width=160pt]{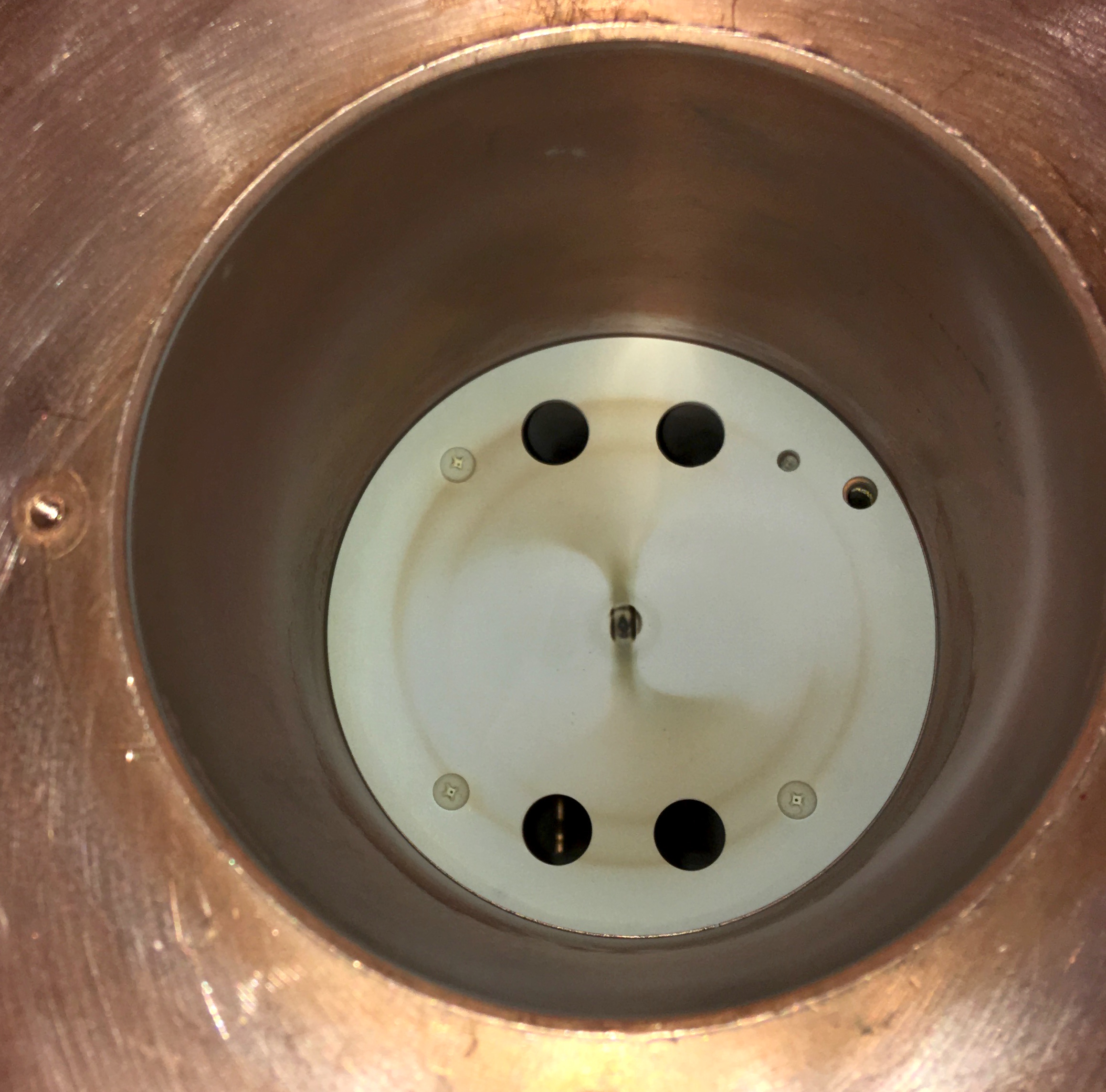}
	   \caption{A damaged BN disk (BN \#2 in Figure~\ref{Structure_IS}) in the LIPAc ion source.}
	   \label{IS_BN2}
	\end{figure}

	The beam from the ion source is injected into the LEBT line where
	many secondary electrons exist. They are produced by collision between the beam and
	residual gas~\cite{footnote1} in the LEBT.
	They are also produced by the collision of the beam with other accelerator components located inside the 
	LEBT. If some of these secondary electrons are attracted by the acceleration voltage,
	they rush into the ion source and damage it. They are therefore called back-streaming
	electrons.
	To prevent back-streaming electron invasion into the ion source, a repeller electrode is
	used (Figure~\ref{Structure_IS} shows a repeller electrode installation).
	Applying a negative voltage to the repeller electrode creates a potential wall
	against the back-streaming electrons~\cite{footnote2}. 
	However, many high-current ion source facilities report damage or countermeasure
	of back-streaming electrons~\cite{LEDA_1997},~\cite{PSI_2011},~\cite{IUCF_2004},~\cite{China_2008}.
	Figure~\ref{IS_BN1} shows a new BN disk (BN \#2 in Figure~\ref{Structure_IS}),  and Figure~\ref{IS_BN2} shows a 
	damaged BN \#2 disk in the LIPAc ion source. The central area is damaged, black colored, and slightly dipped.
	Once a hole is made, the performance is much reduced.  

	One possible scenario of this secondary electron invasion is that the secondary electrons are produced inside
	the extraction system itself. In fact, if the beam divergence becomes large in the extraction system
	due to bad beam optics, the beam can hit the electrodes (such as the puller electrode) of the
	extraction system.
	A large amount of secondary electrons can thus be produced, and they are then attracted to the ion source.

	A second scenario is that the repeller electrode voltage decreases during operation and that the secondary
	electrons located in the LEBT then have sufficient energy to pass the potential of the repeller electrode.
	The drop in the repeller electrode voltage can be due to several reasons.
	For instance, it can be due to a Penning discharge between the repeller and one of the two ground electrodes,
	which is a consequence of electrons accelerated up to several kilovolts and trapped in one or several locations
	in the accelerator column by a combination of electric and magnetic field lines with enough energy to
	ionize the residual gas. 
	The drop in the repeller electrode voltage can be also induced by  collision of the beam with the repeller electrode.
	It has also been observed during LIPAc injector commissioning that a drop in the main high voltage (HV) sometimes
  induced	a drop in the repeller electrode voltage, but more investigations are needed to understand such complex phenomena.
	External disturbances or bad grounding can also induce such a drop in the repeller electrode, as observed during the LIPAc injector commissioning. 

\begin{figure}[!htb]
	   \centering
	   \includegraphics*[width=240pt]{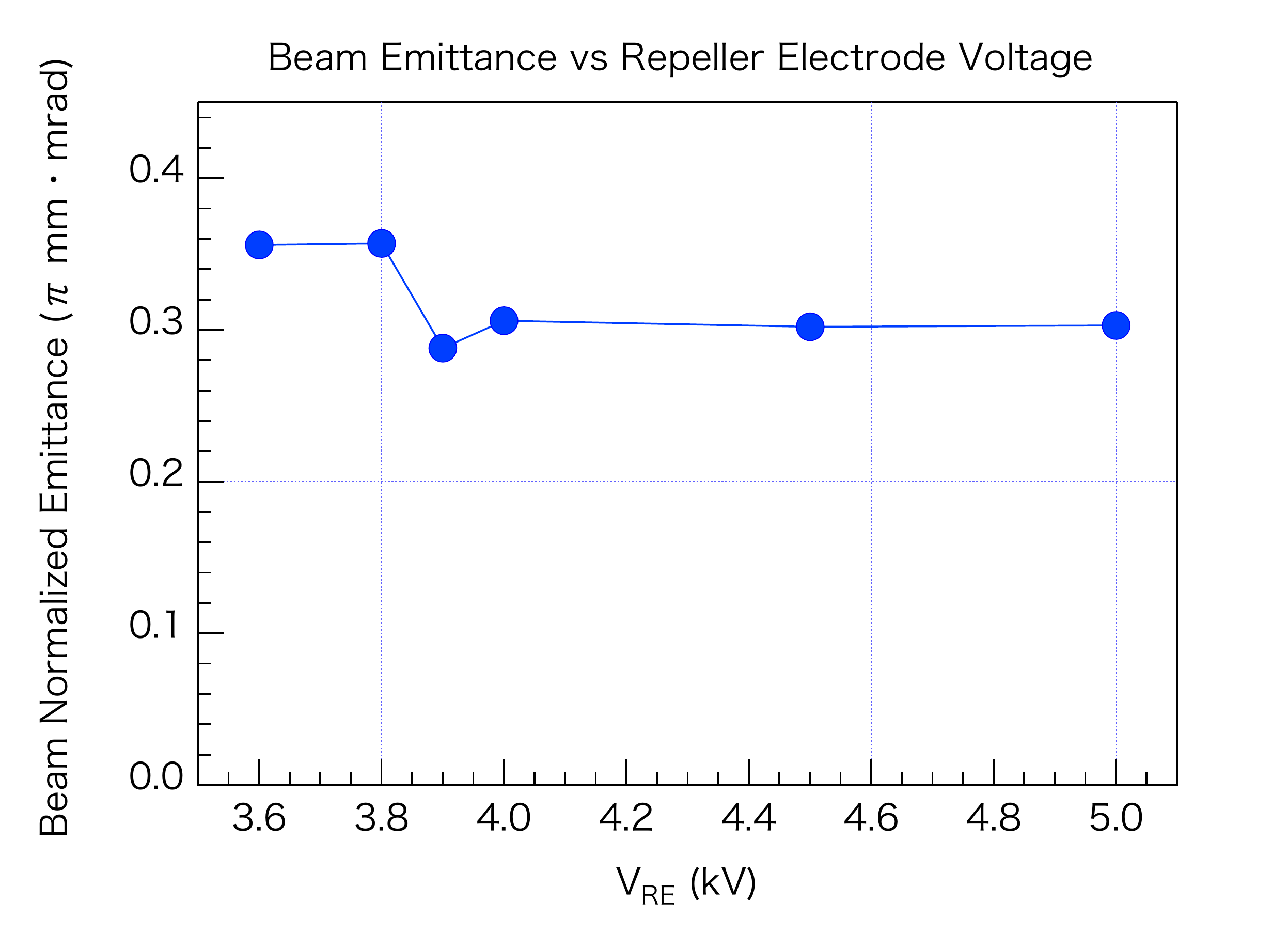}
	   \caption{Emittance measured with a 100~keV deuterium  beam at the LIPAc injector
	   as a function of the repeller electrode voltage to determine the minimum voltage necessary to
	   repel all the secondary electrons from the LEBT.}
	   \label{E_vs_VRE}
	\end{figure}

	To determine the minimum voltage to apply to the repeller electrode to repel all the secondary electrons
	 from the LEBT, measurements of the emittance were performed
	with a 100 keV deuterium beam at the LIPAc injector as a function of the repeller electrode voltage ($\mathrm{{V_{RE}}}$),
	see Figure~\ref{E_vs_VRE}.

	The emittance measurements were performed using an Allison scanner~\cite{Allison} installed just downstream of the extraction system.
	The results show that the emittance increases when the repeller electrode voltage $\mathrm{{V_{RE}}}$ is set below 3.9 kV.
	This observation shows that a minimum voltage of 3.9~kV on the repeller electrode is necessary to
	repel all the secondary electrons. If the voltage is just decreased by 100~V, the energy of the secondary electrons
	(same energy for all electrons) is then enough to pass the potential of the repeller electrode and all these electrons
	come back to the ion source.
	The emittance is thus increased as less secondary electrons are present to contribute to the space charge compensation
	of the beam.
	In addition, it was observed with an oscilloscope that the pulse of total extracted current was becoming unstable and that
	the level of this pulse was increasing for $\mathrm{{V_{RE}}}$ below 3.9~kV. The secondary electrons coming back to the source
	add a positive current to the extracted current and disturb the ion source.
		
	\subsection{Measurement of discharge at the LIPAc injector}
\begin{figure}[!htb]
	   \centering
	   \includegraphics*[width=240pt]{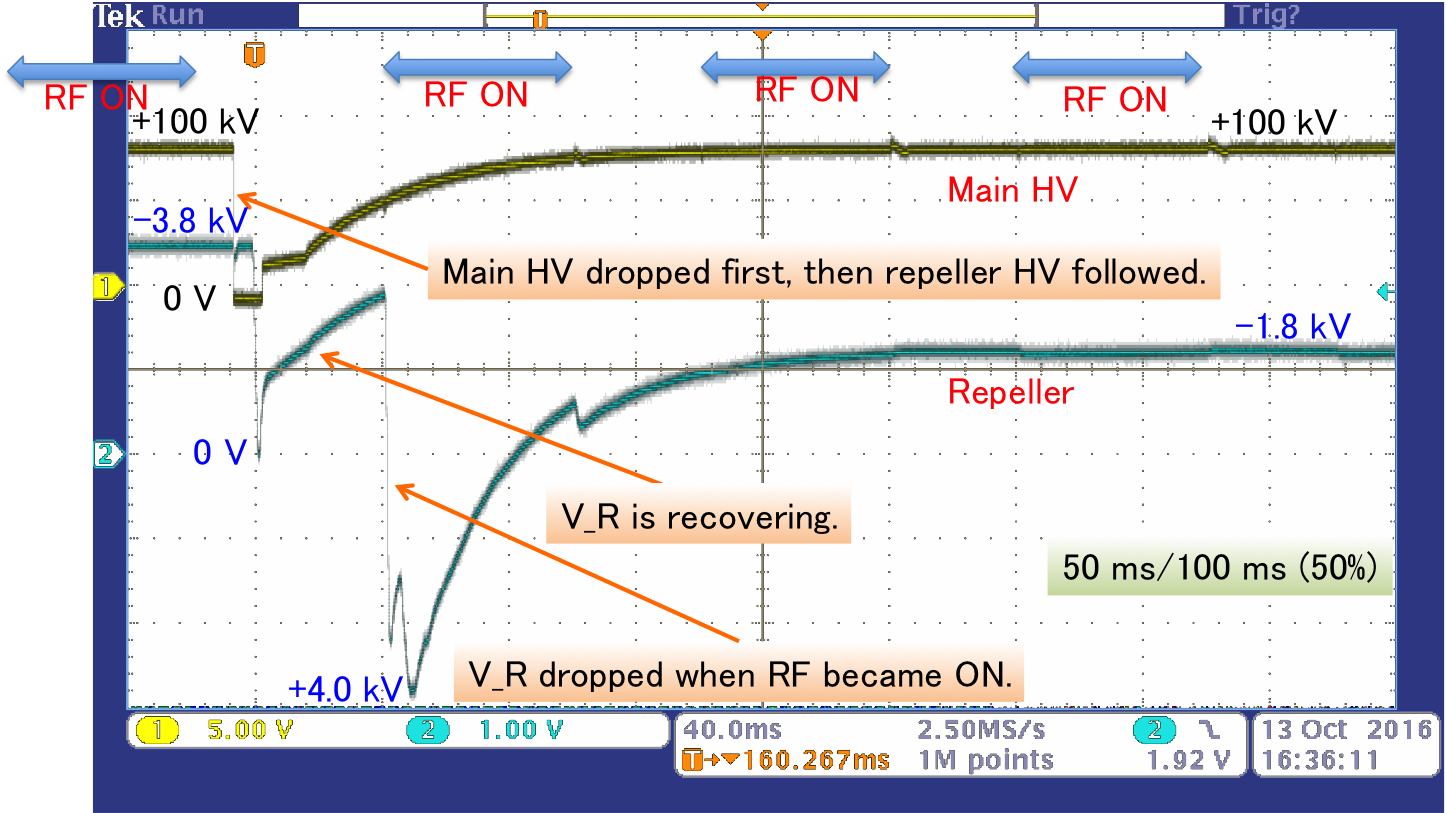}
	   \caption{HVPS monitor outputs when a discharge event occurs.}
	   \label{discharge_scope1}
	\end{figure}

	Figure~\ref{discharge_scope1} shows the main and repeller HV monitor voltage outputs at a typical discharge event at the LIPAc
	ion source.
	A discharge occurred during the time that the RF pulse was off. The main HV dropped to 0 V due to this discharge.
	About 5 ms later, the repeller HV also dropped. Both HV outputs were recovering to the nominal voltage until 
	the next RF pulse was ON. When the next RF pulse was ON, an improper electric field in the extraction system induced 
	a large divergence of the extracted beam and a certain amount of beam hit the repeller electrode.
	This induced a radical change in the repeller electrode potential, which became +4.0~kV. 
	After this event, the repeller HV only partially recovered as it never returned to the nominal voltage
	of $-3.8$~kV but stayed at $-1.8$~kV.
	This voltage was not sufficient to repel the incoming back-streaming electrons as shown previously with the data of
	Figure~\ref{E_vs_VRE}. This state is harmful for the ion source.
	To protect the ion source and its extraction system, the RF pulse should be turned off immediately
	when discharge-related voltage drops are observed.

\section{Implementation of the Discharge Suppression System}

	The objective of the discharge suppression system was to mitigate
	abnormal conditions and downtime of the accelerator.
	The policy of this discharge suppressor system implementation is:
	\begin{itemize}
	\item Implementation using the fast recovery mode of the machine protection system (MPS).
	\item Minimum hardware implementation.
	\item Reuse existing resources as much as possible.
	\item Implementation of core logics by the Experimental Physics and Industrial Control System (EPICS) code (software)~\cite{EPICS}.
	\end{itemize}

\begin{figure}[!htb]
	   \centering
	   \includegraphics*[width=210pt]{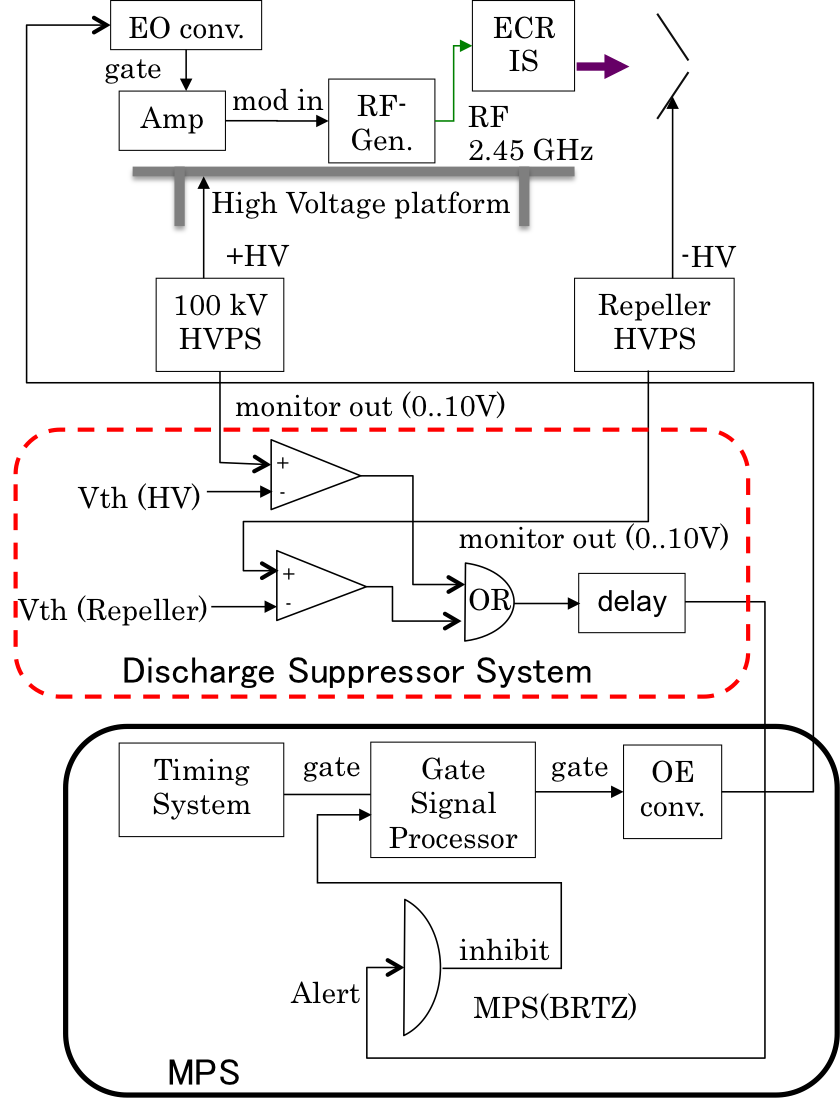}
	   \caption{Discharge suppressor system location in the LIPAc ion source and extraction system.}
	   \label{SparkSuppressor_Impl2}
	\end{figure}

	Figure~\ref{SparkSuppressor_Impl2} shows an overview of this discharge suppressor system.
	The elements of this system are represented inside the dashed rectangle.
	It continuously monitors voltage monitor outputs of the 100~kV high-voltage power supply (HVPS) and repeller HVPS.
	When its comparators detect a voltage drop for at least one monitor input, they send a pre-determined length of alert signal
	to the MPS input. The MPS system \cite{MPS_IPAC11} (below) stops the RF gate signal going to the RF generator of the ion source.

%

\subsection{EPICS code implementation}
	All functionalities (Comparator, OR, and Delay) shown in 
	Figure~\ref{SparkSuppressor_Impl2} 
	except the MPS interface were implemented by the EPICS database (DB) system.
	Comparator, OR, and Event counter were implemented with calc or calcout records.
	Delay functionality was implemented with the function bo record (HIGH field). 
	For further details, see the Appendix.

\section{Results}
	Figure~\ref{tek00013-3} shows typical behavior with  activation of the discharge suppressor system. 
	The blue (ch.2: V\_Rep) line is the repeller HV monitor output voltage, the green (ch.4: V\_HV) line is the 100~kV HV
	monitor output voltage, and the magenta (ch.3: I\_HV) line is the total extraction current of the 100~kV HV power supply
	that corresponds to the beam output current from the ion source. The yellow line (ch1:I\_FC) is the current
	measured with the Faraday cup located in the LEBT~\cite{FaradayCup}.

	At the beginning of the second I\_HV pulse, a 100~kV HV discharge occurred and its voltage dropped to 0~kV.
	The repeller HV (4.5~kV  initially) dropped a few milliseconds later. This discharge suppression system
	prevented RF pulse injection (corresponding to the I\_HV pulse) for 640~ms, and no further HV drops were observed.
	In fact, this time interval of 640~ms allows the main HV and repeller HV to fully 
	recover before the next beam pulse is generated.

	\begin{figure}[!htb]
	   \centering
	   \includegraphics*[width=220pt]{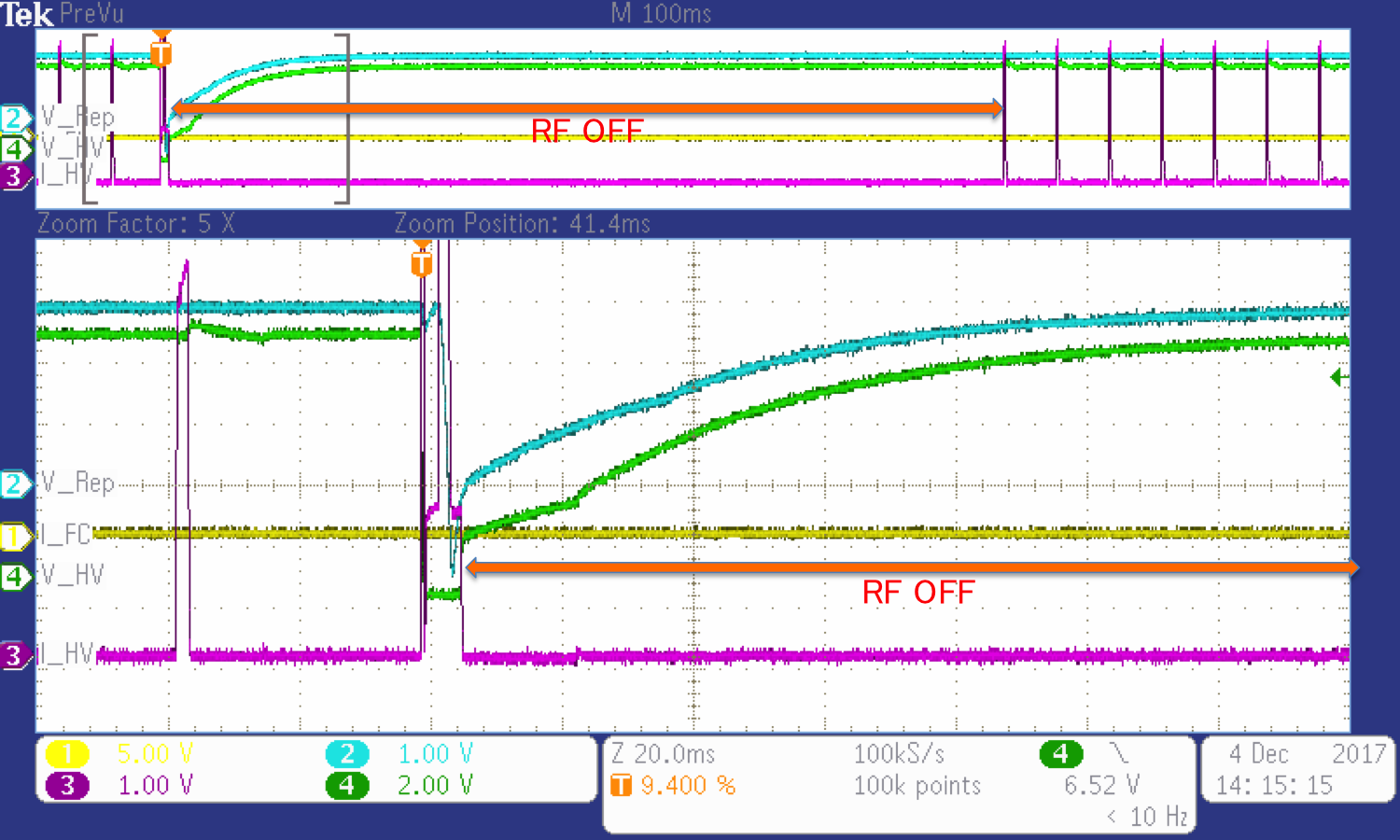}
	   \caption{Typical behavior with  activation of the discharge suppressor system.}
	   \label{tek00013-3}
	\end{figure}

	\begin{figure}[!htb]
	   \centering
	   \includegraphics*[width=220pt]{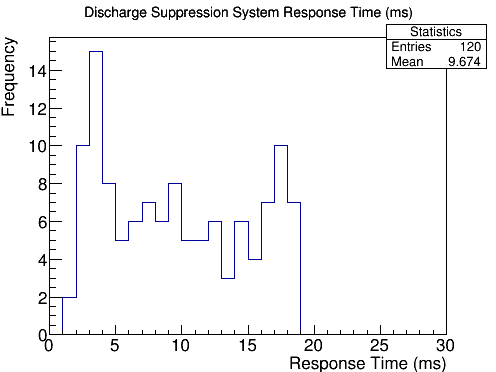}
	   \caption{Discharge suppression system response time distribution.}
	   \label{ss-response-time}
	\end{figure}

	Figure~\ref{ss-response-time} shows response time measurement of the discharge suppression system from the HV monitor drop to the
	MPS alert signal output. These data were obtained by injecting into the system
	a simulated repeller monitor drop signal and by measuring the input and output signal latency with an oscilloscope.
	The response time was randomly distributed from 1~ms to 19~ms. 

	The HV monitor output and repeller monitor output were independently sampled at 50~Hz and processed.
	Therefore, the dominant contribution of these response times was the sampling interval of the EPICS software.
	Generally, a minimum software sampling interval is determined by timer interruption period in the operating system.
	This means 50~Hz is the maximum sampling speed with the current VxWorks~6.9 real-time operating system~\cite{VxWorks} default configuration.

\section{Conclusion}
	Continuous discharge at the repeller electrode can cause invasion of back-streaming electrons
	into the ion source, which is then damaged. 
	To prevent continuous discharge at the ion source and the extraction system, immediate shut-off of the ion source
	RF power is mandatory. 	
	To realize this required functionality, a software-based ion source and beam extraction system suppression system
	has been implemented and tested.
	This system was implemented with an existing VME-based EPICS data acquisition system. All functionalities were implemented with the EPICS DB system.
	It effectively prevented fatal continuous discharge.
	This system is relatively slow for software implementation, but it has demonstrated its ability to suppress continuous discharge efficiently.

\section*{acknowledgments}
	This work was carried out as part of the IFMIF/EVEDA project based on Broader Approach activities.

\section*{APPENDIX:\endgraf DETAILS OF IMPLEMENTATION}
\begin{figure*}[!ph]
    \centering
    \includegraphics*[width=\textwidth]{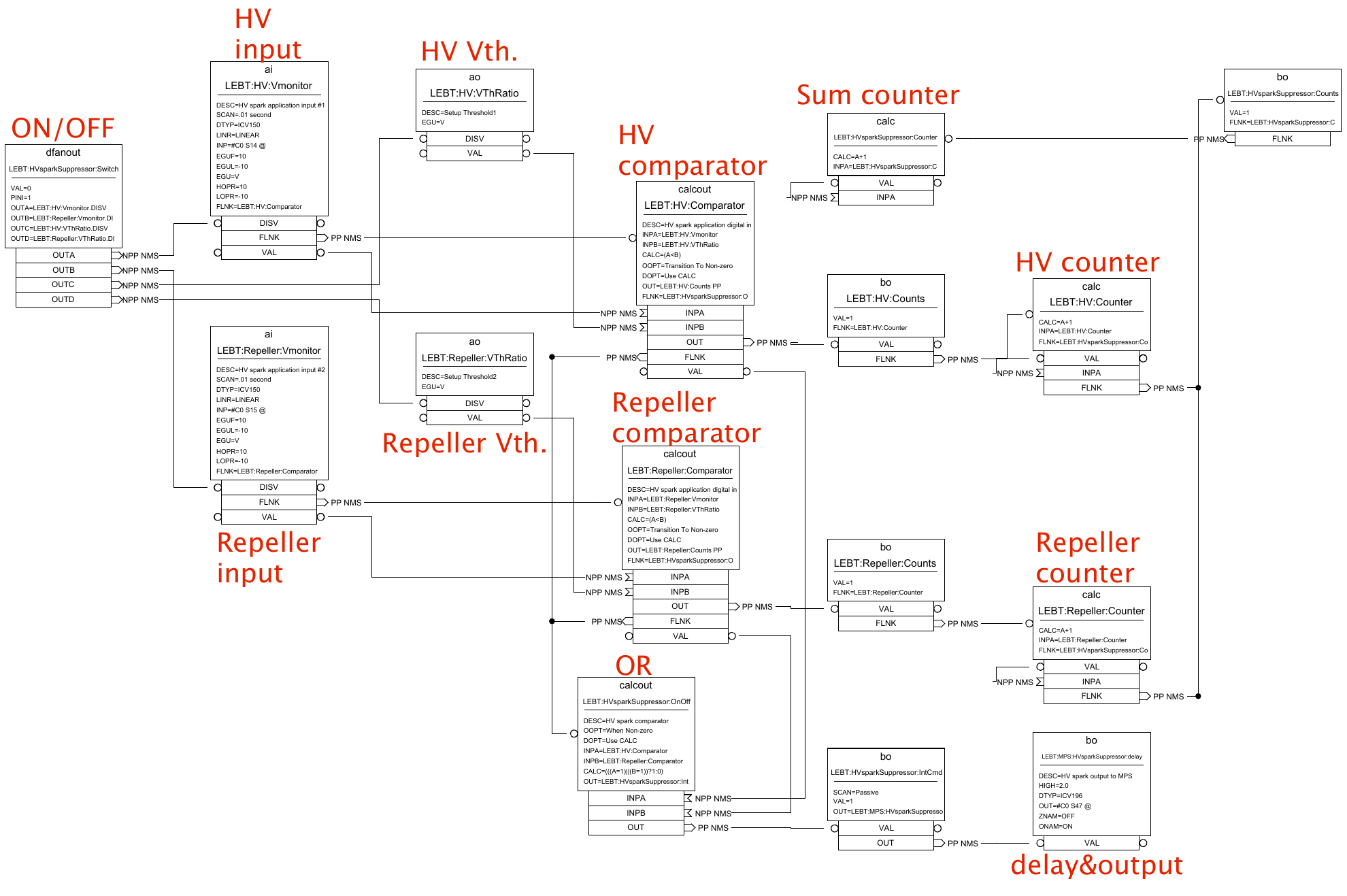}
    \caption{EPICS DB structure diagram.}
    \label{EPICS_DB_Diagram}
\end{figure*}

	As described in the EPICS code section, all functionalities in this system
	except the MPS interface were implemented by the EPICS DB system.
	Figure~\ref{EPICS_DB_Diagram} shows the implemented DB diagram.

\subsubsection{Input}
	ICV150 ADC device support was used, which scanned every 20~ms.
	\begin{verbatim}
# HV monitor input
record(ai, LEBT:HV:Vmonitor){
  field(DESC, "HV spark application input #1")
  field(SCAN, ".02 second")
  field(DTYP, "ICV150")
  field(LINR, "LINEAR")
  field(INP, "#C0 S14 @")
  field(EGUF, "10")
  field(EGUL, "-10")
  field(EGU, "V")
  field(HOPR, "10")
  field(LOPR, "-10")
}
	\end{verbatim}

\subsubsection{Comparator and OR logic}
	The comparator and OR logic were implemented with the calcout record.
	CP input links forced the record routine to process when input variables were changed.
	\begin{verbatim}
# HV comparator
record(calcout, LEBT:HV:Comparator){
  field(DESC, "HV voltage comparator")
  field(INPA, "LEBT:HV:Vmonitor CP")
  field(INPB, "LEBT:HV:VThRatio CP")
  field(CALC,"(A<B)")
  field(OOPT, "Transition To Non-zero")
  field(DOPT, "Use CALC")
  field(OUT, "LEBT:HV:Counts PP")
}

# OR logic
record(calcout, LEBT:HVsparkSuppressor:OnOff){
  field(DESC,"HV spark comparator")
  field(OOPT, "When Non-zero")
  field(DOPT, "Use CALC")
  field(INPA,"LEBT:HV:Comparator CP")
  field(INPB,"LEBT:Repeller:Comparator CP")
  field(CALC,"((A=1)||(B=1))")
  field(OUT, "LEBT:HVsparkSuppressor:IntCmd PP")
}
	\end{verbatim}

\subsubsection{Delay and output}
	When the VAL field was set to 1, if the HIGH field was greater than 0, the record routine
	process was reset to 0 after HIGH seconds~\cite{EPICS-WIKI-DELAY}.
	The ICV196 digital output device support was used to output the digital signal.
	\begin{verbatim}
# delay & output
record(bo, LEBT:MPS:HVsparkSuppressor:delay){
  field(DESC, "HV spark output to MPS")
  field(HIGH, "0.5")
  field(DTYP, "ICV196")
  field(OUT, "#C0 S47 @")
  field(ZNAM, "OFF")
  field(ONAM, "ON")
}
	\end{verbatim}

\begin{thebibliography}{99}
\bibitem{footnote1} Typical pressure in the LEBT is a few $\mathrm{\times 10^{-3}}$ Pa.
\bibitem{footnote2} Back-streaming electrons generated by beam collision at the extraction system cannot be
	protected by the repeller electrode potential.
\bibitem{FaradayCup} No Faraday Cup current was observed at that time since the emittance meter was inserted at this time.
\end{thebibliography}
\end{document}